\documentclass[runningheads]{llncs}
\usepackage[T1]{fontenc}
\usepackage{graphicx}
\usepackage{array}
\usepackage{booktabs}
\usepackage{subfigure}
\usepackage{parskip}

\usepackage{makecell}
\usepackage{hyperref}
\usepackage{multirow}
\begin{document}
\title{Investigating Gender Euphoria and Dysphoria on TikTok: Characterization and Comparison}
\author{SJ Dillon\inst{1}$^*$ \and
% Yueqing Liang\inst{2}\textsuperscript{\scriptsize \textasteriskcentered} \and
Yueqing Liang\inst{2}$^*$ \and
H. Russell Bernard\inst{3} \and 
Kai Shu\inst{1}}
\authorrunning{SJ et al.}
\institute{Emory University, Atlanta GA 30322, USA \and
Illinois Institute of Technology, Chicago IL 60616, USA \and
Arizona State University, Tempe AZ 85281, USA\\
\email{sarah.dillon@emory.edu, yliang40@hawk.iit.edu, asuruss@asu.edu, kai.shu@emory.edu}}
\maketitle              % typeset the header of the contribution
%
% \renewcommand{\thefootnote}{\fnsymbol{footnote}}
% \footnotetext[1]{Equal contribution.}
\setcounter{footnote}{1}
\def\thefootnote{*}\footnotetext{Equal contribution.}\def\thefootnote{\arabic{footnote}}

\begin{abstract}
% With the emergence of short video-sharing platforms, engagement with social media sites devoted to opinion and knowledge dissemination has rapidly increased. Among the short video platforms, TikTok is one of the most popular globally and has become the platform of choice for transgender and nonbinary individuals, who have formed a large community to mobilize personal experience and exchange information. The knowledge produced in online spaces can influence the ways in which people understand and experience their own gender and transitions, as they hear about others and weigh that experiential and medical knowledge against their own. This paper extends current research and past interview methods on gender euphoria and gender dysphoria to analyze what and how online communities on TikTok discuss these two types of gender experiences. Our findings indicate that gender euphoria and gender dysphoria are differently described in online TikTok spaces. These findings indicate that there are wide similarities in the words used to describe gender dysphoria as well as gender euphoria in both the comments of videos and content creators’ hashtags. Finally, our results show that gender euphoria is described in more similar terms between transfeminine and transmasculine experiences than gender dysphoria, which appears to be more differentiated by gendering experience and transition goals. We hope this paper can provide insights for future research on understanding transgender and nonbinary individuals in online communities.
With the emergence of short video-sharing platforms, engagement with social media sites devoted to opinion and knowledge dissemination has rapidly increased. Among these platforms, TikTok is one of the most popular globally and has become the platform of choice for transgender and nonbinary individuals, who have formed a large community to mobilize personal experience and exchange information. The knowledge produced in online spaces can influence the ways in which people understand and experience their own gender and transitions, as they hear about others and weigh experiential and medical knowledge against their own. This paper extends current research and past interview methods on gender euphoria and gender dysphoria to analyze what and how online communities on TikTok discuss these two types of gender experiences. Our findings indicate that gender euphoria and gender dysphoria are differently described in online TikTok spaces. These findings indicate similarities in the words used to describe gender dysphoria as well as gender euphoria in both the comments of videos and content creators’ hashtags. Finally, our results show that gender euphoria is described in more similar terms between transfeminine and transmasculine experiences than gender dysphoria, which appears to be more differentiated by gendering experience and transition goals. We hope this paper can provide insights for future research on understanding transgender and nonbinary individuals in online communities.

\end{abstract}
\section{Introduction}

With the emergence of short video-sharing platforms, engagement with social media sites devoted to opinion and knowledge dissemination has rapidly increased ~\cite{mackinnon2021examining}. These videos are easy to access and understand, and can be quickly disseminated, driving community formation in online spaces. TikTok is one of the most popular of these platforms globally, with nearly 45.1 million daily active users
\footnote{\href{https://www.demandsage.com/tiktok-user-statistics}{https://www.demandsage.com/tiktok-user-statistics}}
, and has become the platform of choice for transgender and nonbinary individuals (TNB), who have formed a large community exchanging information and sharing knowledge~\cite{mackinnon2021examining}. Extensive online ethnography indicates that this knowledge comes from diverse sources, including medical experts, activists, and influencers, as well as from community members who share information about accessing medical services, personal experiences, what members of the community call “trans joy,” traumatic encounters, and intimate changes in their bodies. The knowledge produced in online spaces can influence the ways in which people understand and experience their own gender and transitions, as they hear about others and weigh that personal knowledge against their own. Therefore, in this paper, we extend the current research on gender euphoria and gender dysphoria to reveal the ways in which online communities on TikTok discuss these two differing but related types of gender experience.

Dysphoria is in many ways considered foundational to the recognition of transgender people today, although this is a problematic definition~\cite{jacobsen2022counts}. Dysphoria is often defined as the distress that begets trans identity, more specifically as “clinically significant distress or impairment related to a strong desire to be of another gender, which may include the desire to change primary and/or secondary sex characteristics”\cite{turban2020what}. Gender euphoria, on the other hand, is often considered gender dysphoria’s inverse or corollary, and is defined as the heightened joy transgender people experience when their gender is recognized and affirmed~\cite{austin2022gender,beischel2022little,jacobsen2022moving,rachlin2018medical}. Gender dysphoria is largely a medical term that grew from the efforts of trans activists to increase access to medical treatment in conversations with medical practitioners grappling with trans identity in the 1990s
% \footnote{Some sources, such as ~\cite{johnson2019rejecting}, claim it was a medical term, while others, such as ~\cite{konnelly2022transmedicalism} and ~\cite{valentine2007imagining}, argue it was coined by gender-diverse people. Its most standardized definition is from the ~\cite{american10diagnostic}, defining gender dysphoria as “distress that may accompany the incongruence between one's experienced or expressed gender and one's assigned gender.”}
. In contrast, gender euphoria rose in use primarily in the past decade or so due to the efforts by community members to have the positive aspects of their gender transition and attendant experiences recognized. Gender dysphoria still largely serves as the label that can officially produce trans identity in medical spaces, because gender-affirming care (i.e., medical transition) is not available unless gender dysphoria has been diagnosed~\cite{shuster2021trans}, although this does not capture the nuance of many trans and gender nonbinary people’s experiences.

\begin{figure}[t]
    \centering
    \subfigure[TikTok video]{
        \includegraphics[width=0.3\linewidth]{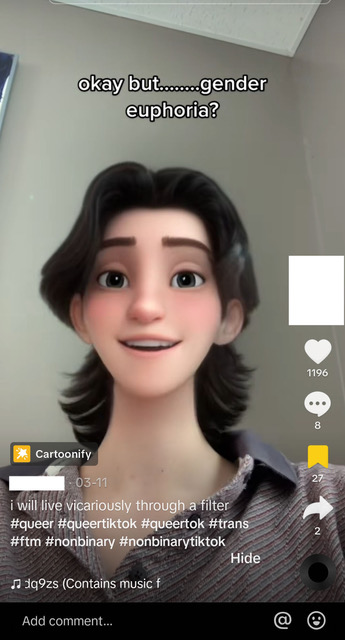}
        \label{fig:figure_1_1}
    }
    \subfigure[Video comments]{
	      \includegraphics[width=0.3\linewidth]{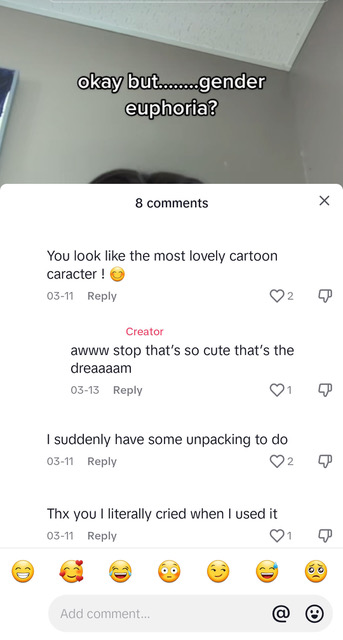}
        \label{fig:figure_1_2}
    }
    \caption{An example of a TikTok video.}
    \label{fig:figure_1}
\end{figure}

TikTok is a video-sharing platform with three main components: the video itself, the comments, and the hashtags that content creators apply to their videos. The video includes a combination of sounds (the auditory backdrop of the video) and the visual field in which the content creator appears. Videos are organized and searchable by the hashtags that content creators apply directly to their videos and are visible to viewers of the video within the TikTok application. The comments section is the space in which TikTok users can interact with the creators of the videos and with each other, to give feedback, have conversations, make jokes, and be in community. Figure~\ref{fig:figure_1} illustrates a typical TikTok discussing gender euphoria. This is an actual video that has been deidentified and displays an individual using an anime filter, which obscures their face. Figure 1a illustrates what a video looks like when accessing TikTok, as well as the video description of a TikTok with its hashtags. Figure 1b illustrates the comments under this particular TikTok, and illustrates how users and content creators interact with another in conversation. We focus our analysis on the comments sections and the hashtags applied to TikToks such as this to investigate our central questions~\cite{mackinnon2021examining}. This research will analyze how transgender communities on TikTok define, use, and experience gender euphoria and gender dysphoria to support data-driven and nuanced consideration of these terms across a wide bandwidth of people. This research ultimately works to respect that transgender people are the experts of their own diverse experiences \cite{solutions2022deeper}. 

Here we investigate: (1) how communities of transgender people organize their experiences of gender dysphoria and gender euphoria using those algorithmic structures, and what kinds of emotional weight those terms carry for both content creators and commentors; and (2) how the organization of transmasculine and transfeminine\footnote{Transfeminine and transmasculine are terms used to describe transgender experiences. Transmasculine refers to people who are transitioning to a more masculine gender and/or presentation, including trans men, FTM (female-to-male) transsexuals, transmasculine nonbinary people, and so on. Transfeminine refers to people who are transitioning to a more feminine gender and/or presentation, including trans women, MTF (male-to-female) transsexuals, transfeminine nonbinary people, and the like. Our data do not currently get at the experiences of transgender nonbinary people who do not identify as transmasculine or transfeminine.} communities on TikTok intersect or do not around the topics of gender dysphoria and gender euphoria. According to the above, we could ask the following research questions (RQs):
% These questions guide us to ask questions from multiple perspectives:

\begin{itemize}
    \item \textbf{RQ1}: What types of words are used and how are those words correlated for TikTok comment sections relating to the gender dysphoria and gender euphoria hashtags?
    \item \textbf{RQ2}: What types of words are used and how are those words correlated for TikTok descriptions relating to the gender dysphoria and gender euphoria hashtags?
    \item \textbf{RQ3}: What sentiments are associated with gender euphoria and gender dysphoria hashtags?
    \item \textbf{RQ4}: What types of topics are discussed in the gender dysphoria and gender euphoria TikTok online space?
\end{itemize}

\section{Dataset}

To explore how people express their gender euphoria and dysphoria in TikTok videos, we scraped data from TikTok\footnote{\href{https://www.tiktok.com/}{https://www.tiktok.com/}} using the TikTok API tool\footnote{\href{https://davidteather.github.io/TikTok-Api/docs/TikTokApi.html}{https://davidteather.github.io/TikTok-Api/}}. We extracted videos marked with the hashtag \#gendereuphoria or \#genderdysphoria, and then collected the video descriptions and comments for each video. Our dataset comprises 116 and 110 videos for each of the two main hashtags and includes 1 description and 100 comments for each video. The statistics of the dataset are in Table~\ref{tab:data_stats}.

\begin{table}[t!] \centering
    \caption{Statistics of the collected data.}
    \label{tab:data_stats}
    \begin{tabular}{l|l|l|l}
      \toprule
      \textbf{Hashtag} & \makecell[l]{\textbf{\#Videos}} & \makecell[l]{\textbf{\#Video }\\\textbf{description}} & \textbf{\#Cmts.}\\
      \hline
      \textit{\#Gendereuphoria}  & 116 & 116 & 12,647\\
      \hline
      \textit{\#Genderdysphoria} & 110 & 110 & 12,033\\
      \hline
    \end{tabular}
\end{table}

% \begin{table}[t!] \centering
%     \caption{Statistics of the collected data.}
%     \label{tab:data_stats}
%     \begin{tabular}{l|l|l}
%       \toprule
%       \textbf{Hashtag} & \makecell[l]{\textbf{\#Videos}} & \textbf{\#Cmts.}\\
%       \hline
%       \textit{\#Gendereuphoria}  & 116 & 12,647\\
%       \hline
%       \textit{\#Genderdysphoria} & 110 & 12,033\\
%       \hline
%     \end{tabular}
% \end{table}

\begin{figure}[!ht] %htbp
    \centering
    \subfigure[\#Gendereuphoria]{
        \includegraphics[width=0.47\linewidth]{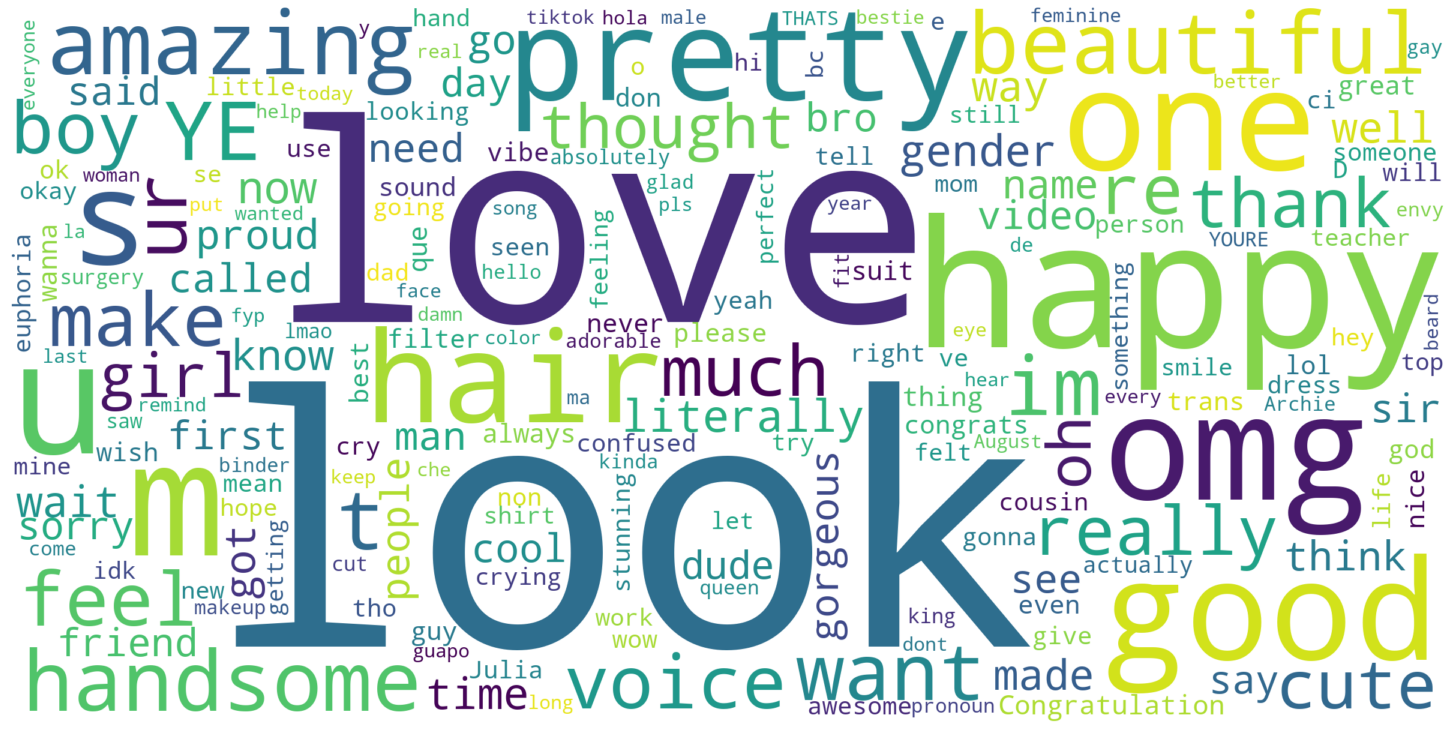}
        \label{fig:wordCloud_cmmt_eu}
    }
    \subfigure[\#Genderdysphoria]{
	      \includegraphics[width=0.47\linewidth]{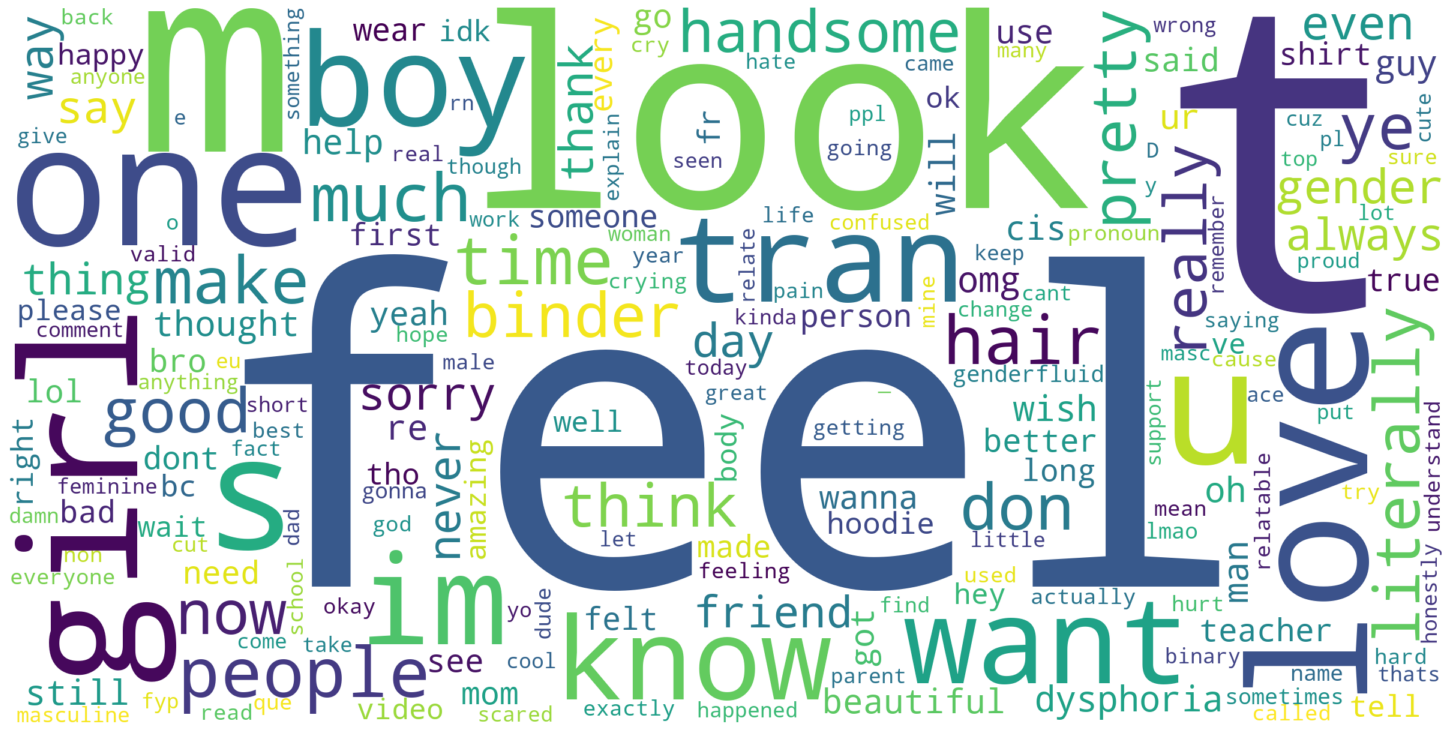}
        \label{fig:wordCloud_cmmt_dys}
    }
    \caption{Word cloud of comments for different hashtags.}
    \label{fig:wordCloud_cmmt}
\end{figure}

\section{Analysis}
\label{sec:analysis}

We group the data into: (1) video descriptions for \#gendereuphoria; (2) comments for \#gendereuphoria; (3) video descriptions for \#genderdysphoria; and (4) comments for \#genderdysphoria and analyze the data with \textsc{word clouds}, \textsc{semantic network analysis}, \textsc{Spearman’s rank-order correlation}, \textsc{sentiment analysis}, and \textsc{Latent Dirichlet Allocation (LDA) analysis}.

\begin{center}
\fbox{\parbox[c]{.9\linewidth}{\textbf{RQ1: What types of words are used and how are those words correlated with each other for TikTok comments about gender dysphoria and gender euphoria hashtags?}
}}
\end{center}

% \subsubsection{Word Cloud}
\label{sec:wordCloud_cmmt}

To answer \textbf{RQ1}, we first use \textsc{Word Clouds} to visualize which words people usually use to describe these two hashtags. A word cloud is a visualization of word frequency in a set of texts. Figure~\ref{fig:wordCloud_cmmt} shows the words TikTokers typically use in describing gender euphoria or gender dysphoria. 

\emph{There are wide similarities in the words used to describe gender dysphoria and gender euphoria in the comments on videos.} The comments often focus on visual aspects of the videos, with common terms such as “love”, “look”, and “hair” appearing in both word clouds. In the gender euphoria word cloud, positive terms like “good”, “amazing”, “happy”, “pretty”, “handsome”, and “beautiful” are prominent. In contrast, the gender dysphoria word cloud features more ambiguous terms like “one”, “want”, “girl”, and “boy”, with “look” and “feel” being the largest, highlighting the importance of visual and affective responses in the comments.

\begin{figure*}[!ht]
    \centering
    \subfigure[\#Gendereuphoria]{
        \includegraphics[width=0.45\linewidth]{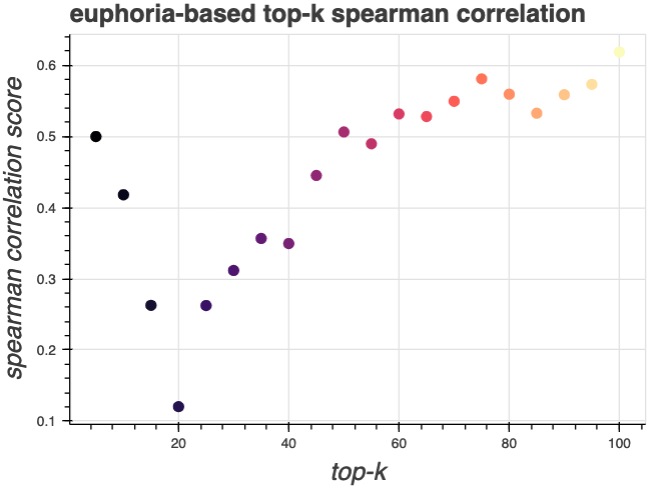}
        \label{fig:spearman_cmmt_eu}
    }
    \subfigure[\#Genderdysphoria]{
	      \includegraphics[width=0.45\linewidth]{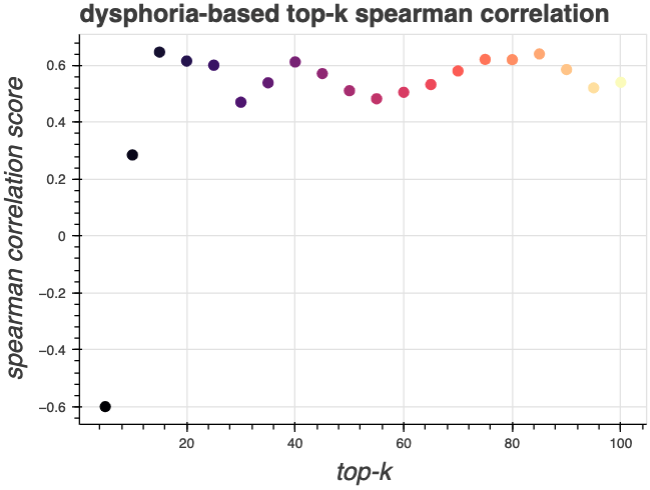}
        \label{fig:spearman_cmmt_dys}
    }
    \caption{Spearman’s correlation between the top words of comments for different hashtags, by changing the $k$.}
    \label{fig:spearman_cmmt}
\end{figure*}

% \subsubsection{Spearman’s Rank Order Coefficient}
\label{sec:spearman_cmmt}

While word clouds provide an overall view of word frequency differences, they lack quantitative detail. Therefore, we use \textsc{Spearman’s Rank-order Coefficient} to numerically distinguish word frequencies. We calculated Spearman’s coefficient by varying $k$ from 5 to 100 to observe changes in correlation between the top-$k$ words in comments for each hashtag. 
Figure~\ref{fig:spearman_cmmt_eu} and Figure~\ref{fig:spearman_cmmt_dys} show that the top-$k$ words for gender euphoria and gender dysphoria comments generally have similar ranks. However, Figure~\ref{fig:spearman_cmmt_eu} demonstrates that rank similarity varies with $k$, decreasing until $k$=20, then stabilizing. This suggests a latent discrepancy in how TikTok commenters view gender euphoria and dysphoria, particularly in the top 20 words. In contrast, the trend in Figure~\ref{fig:spearman_cmmt_dys} is more stable. This difference likely arises because commenters exhibit more excitement about gender euphoria but avoid disparaging experiences of gender dysphoria. Sentiment analysis (Table~\ref{tab:sentiment}) supports this interpretation, showing positive attitudes towards gender euphoria and neutral sentiments towards gender dysphoria. 

% \subsubsection{Semantic Network Analysis}
\label{sec:semantic_cmmt}

\begin{figure*}[ht!]
    \centering
    \subfigure[\#Gendereuphoria]{
        \includegraphics[width=0.47\linewidth]{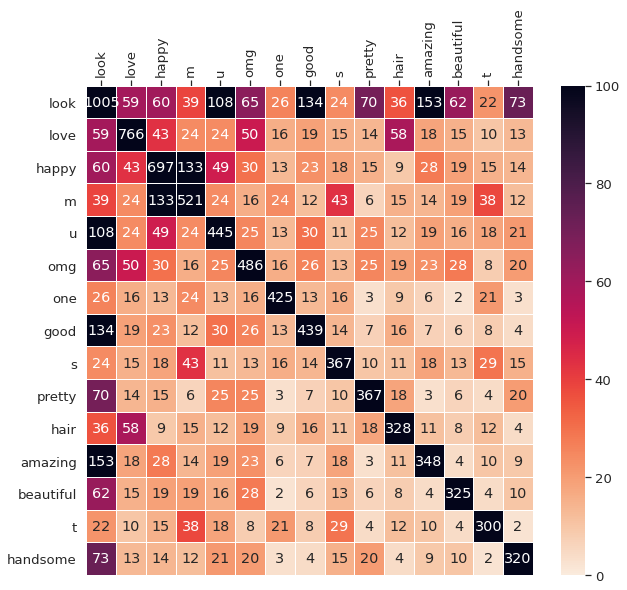}
        \label{fig:semantic_cmmt_eu}
    }
    \subfigure[\#Genderdysphoria]{
	      \includegraphics[width=0.47\linewidth]{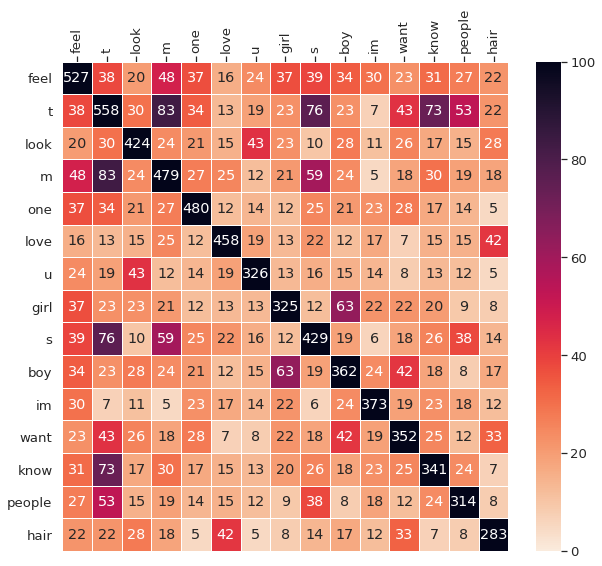}
        \label{fig:semantic_cmmt_dys}
    }
    \caption{Semantic Network Analysis for word-pairs in comments of different hashtags.}
    \label{fig:semantic_cmmt}
\end{figure*}

We also conducted \textsc{Semantic Network Analysis} to examine word co-occurrence patterns. Analyzing the top-15 words for both hashtags, the heat maps in Figure~\ref{fig:semantic_cmmt} reveal relationalities among words.
In \#gendereuphoria comments, words like “look” and “love” co-occur in 59 comments, while “look” and “good” co-occur in 134 comments, indicating a strong correlation between “look” and “good”. People often use phrases like “look good” or “look amazing” to describe gender euphoria. “Hair” is also a significant term in these comments. 
For \#genderdysphoria comments, the focus is more on feelings, with words like “feel” being prominent. The term “t” (testosterone) suggests a dominance of transmasculine conversations. Words like “m” (masculine) and “know” also show high correlation frequencies.

\begin{center}
\fbox{\parbox[c]{.9\linewidth}{\textbf{RQ2: What types of words are used, and how are those words correlated with each other for TikTok descriptions under the gender dysphoria and gender euphoria hashtags?}
}}
\end{center}

% \subsubsection{Word Cloud}
\label{sec:wordCloud_desc}

To explore the words that video creators use to title their videos (\textbf{RQ2}), we apply the same methods as in RQ1 to TikTok video descriptions under the gender dysphoria and gender euphoria hashtags.

\begin{figure}[ht!]
    \centering
    \subfigure[\#Gendereuphoria]{
        \includegraphics[width=0.47\linewidth]{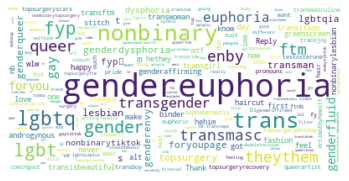}
        \label{fig:wordCloud_desc_eu}
    }
    \subfigure[\#Genderdysphoria]{
	      \includegraphics[width=0.47\linewidth]{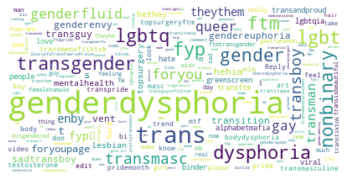}
        \label{fig:wordCloud_desc_dys}
    }
    \caption{Word cloud of video descriptions for the gender euphoria and gender dysphoria hashtags.}
    \label{fig:wordCloud_desc}
\end{figure}

Figure~\ref{fig:wordCloud_desc_eu} and Figure~\ref{fig:wordCloud_desc_dys} show that \emph{words used under both hashtags show broad similarity.} Common terms include “trans”, “LGBTQ”, “nonbinary”, “for you page”, “fem”, “genderenvy”, “they/them”, “enby”, and “transmac”. Transmasculine terms, such as “masc”, “trans boy”, “transmasc”, “transman”, “transguy”, and “top surgery”, are more common than transfeminine terms like “fem”. This could be due to our sample or a higher number of TikTok videos focusing on transmasculine experiences.
As expected, “gender dysphoria” and “dysphoria” are more common in the dysphoria sample, and “gender euphoria” and “euphoria” in the euphoria sample. Interestingly, “gender euphoria” appears in the dysphoria cloud, but “gender dysphoria” does not appear in the euphoria cloud. This suggests a recognition of euphoria during dysphoria but not vice versa.

\begin{figure*}[t]
    \centering
    \subfigure[\#Gendereuphoria]{
        \includegraphics[width=0.45\linewidth]{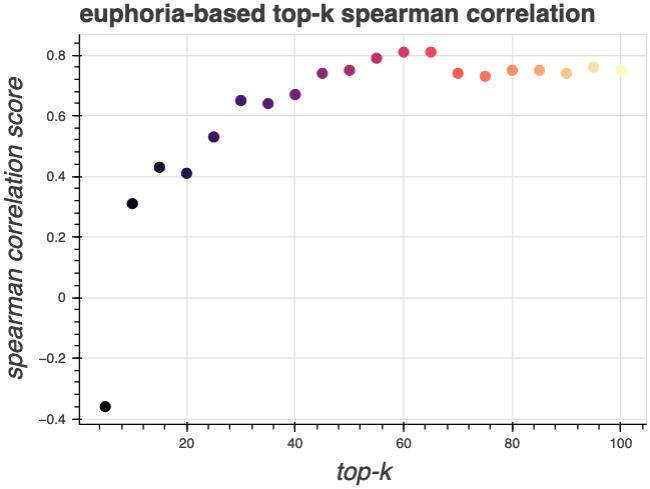}
        \label{fig:spearman_desc_eu}
    }
    \subfigure[\#Genderdysphoria]{
	      \includegraphics[width=0.45\linewidth]{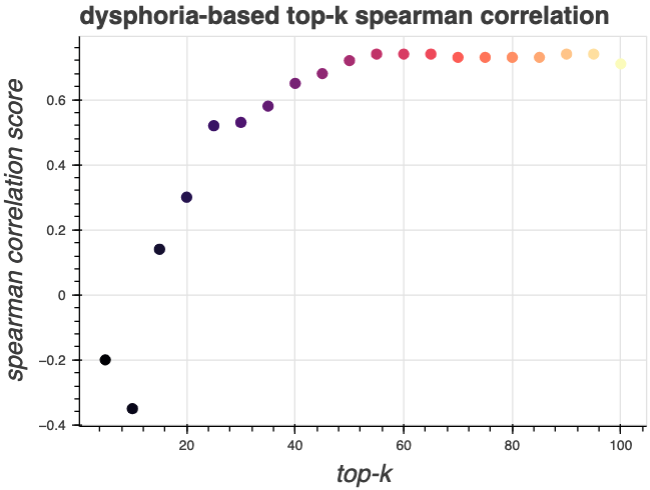}
        \label{fig:spearman_desc_dys}
    }
    \caption{Spearman’s correlation between the top words of video descriptions for different hashtags, by changing the $k$.}
    \label{fig:spearman_desc}
\end{figure*}

\label{sec:spearman_desc}

Spearman’s coefficient for video descriptions shows a positive correlation with $k$ (see Figure~\ref{fig:spearman_desc}). When $k$ is small, the top-$k$ words in one dataset have very different ranks in the other dataset. As $k$ increases, Spearman’s correlation score also increases, indicating that on a larger scale, the difference diminishes. This shows broad similarities in how creators title TikToks about gender dysphoria and gender euphoria.

Figure~\ref{fig:spearman_desc_eu} and~\ref{fig:spearman_desc_dys} demonstrate that while the top-$k$ words for gender euphoria and gender dysphoria are different at smaller $k$ values, the correlation increases with larger $k$ values. This indicates that although specific terms differ initially, broader similarities emerge as more terms are considered, reflecting common themes in how creators title their TikToks about gender experiences. The data suggest that the frequency range in video descriptions for both hashtags tends to converge, highlighting the interconnected nature of gender-related content on TikTok.

\begin{center}
\fbox{\parbox[c]{.9\linewidth}{\textbf{RQ3: What sentiments are associated with gender euphoria and gender dysphoria hashtags?}
}}
\end{center}

% \subsubsection{Sentiment Analysis}
\label{sec:sentiment}

To investigate \textbf{RQ3}, we conducted \textsc{sentiment analysis} with the Vader\footnote{\href{https://github.com/cjhutto/vaderSentiment}{https://github.com/cjhutto/vaderSentiment}} sentiment analysis tool, which is specifically designed for online social networks, to estimate whether the opinions expressed in a given text are positive, negative, or neutral. We ran a sentiment analysis for each comment/video description and report the mean and standard deviation of sentiment scores for each dataset. Scores $<-0.05$ imply negative sentiment;  scores between -0.05 and 0.05 imply a neutral sentiment; and scores $>0.05$ imply a positive sentiment. The results are shown in Table~\ref{tab:sentiment}. From the mean of each dataset, we observe the overall sentiment of users about gender euphoria and gender dysphoria, as expressed in their comments and video descriptions. In contrast to the diversity in the comments, the descriptions for the two hashtags were quite similar. Creators of videos experiencing euphoria and dysphoria themselves expressed little difference when describing the two experiences, at least in the descriptions for the videos. This contrasts with our findings in Spearman’s $k$ correlation analysis of comments for those same videos. We conclude that creators and viewers use different expressions of language when focusing on gender euphoria and gender dysphoria. However, this difference could be because video descriptions serve as branding or searchable signposts by which users can find content. In contrast, comment sections can allow for more conversation about a topic. 

In our sentiment analysis, the positive associations with gender euphoria and the more neutral emotional weight of gender dysphoria, more negative than that of gender euphoria, are clear in both the comment sections and content creator descriptions for TikToks tagged with \#genderdysphoria and \#gendereuphoria. The sentiment means of the comment section for both gender dysphoria and gender euphoria are much more positive than those for the hashtags of the content creators. This indicates that the emotional weight of TikToks relating to gender dysphoria was much more negative and the positive implications of gender euphoria were significantly less positive for content creators than for commenters. Though comment sections for TikToks related to gender euphoria were statistically much more positive than for those related to gender dysphoria, comment sections related to gender dysphoria remained positive in sentiment and were significantly more positive than were the sentiment analysis for the content creators’ descriptions for those videos. This may indicate commenters’ attempt to allay the severely negative sentiment of content creators through uplifting comments and positive feedback on their videos relating to their gender dysphoria. Likewise, commenters displayed statistically more positive sentiment toward gender euphoria than did content creators, perhaps again seeking to support and affirm the gender euphoria of content creators. 

\begin{table}[ht] \centering
    \caption{Sentiment analysis of comments and descriptions for the gender euphoria and gender dysphoria hashtags. Mean$<0.05$: neutral, mean $>0.05$: positive.}
    \label{tab:sentiment}
    \begin{tabular}{l|l|l|l}
    \toprule 
    & \textbf{ Hashtag } & \textbf{ Mean } & \textbf{ Std } \\
    \hline \multirow{2}{*}{Comments} & \textit{ \#Genderdysphoria } & 0.171 & 0.421 \\
    \cline { 2 - 4 } & \textit{ \#Gendereuphoria } & 0.333 & 0.422 \\
    \hline \multirow{2}{*}{Descriptions} & \textit{ \#Genderdysphoria } & 0.040 & 0.387 \\
    \cline { 2 - 4 } & \textit{ \#Gendereuphoria } & 0.266 & 0.413 \\
    \hline
    \end{tabular}
\end{table}

% Table layout in the same line using minipages
% \begin{table}[t]
%     \centering
%     \begin{minipage}{0.45\textwidth}
%         \centering
%         \caption{Statistics of the collected data. Each video has one video description. Therefore, the number of video descriptions is the same as the number of videos.}
%         \label{tab:data_stats}
%         \begin{tabular}{l|l|l}
%             \toprule
%             \textbf{Hashtag} & \makecell[l]{\textbf{\#Videos}} & \textbf{\#Cmts.}\\
%             \hline
%             \textit{\#Gendereuphoria}  & 116 & 12,647\\
%             \hline
%             \textit{\#Genderdysphoria} & 110 & 12,033\\
%             \hline
%         \end{tabular}
%     \end{minipage}
%     \hfill
%     \begin{minipage}{0.45\textwidth}
%         \centering
%         \caption{Sentiment analysis of comments and descriptions for the gender euphoria and gender dysphoria hashtags. Mean$<$0.05: neutral, mean $>$0.05: positive. Cmts. stands for comments, Descs. stands for descriptions. }
%         \label{tab:sentiment}
%         \begin{tabular}{l|l|l|l}
%             \toprule 
%             & \textbf{Hashtag} & \textbf{Mean} & \textbf{Std} \\
%             \hline \multirow{2}{*}{Cmts.} & \textit{\#Genderdysphoria} & 0.171 & 0.421 \\
%             \cline { 2 - 4 } & \textit{\#Gendereuphoria} & 0.333 & 0.422 \\
%             \hline \multirow{2}{*}{Descs.} & \textit{\#Genderdysphoria} & 0.040 & 0.387 \\
%             \cline { 2 - 4 } & \textit{\#Gendereuphoria} & 0.266 & 0.413 \\
%             \hline
%         \end{tabular}
%     \end{minipage}
% \end{table}

\begin{center}
\fbox{\parbox[c]{.9\linewidth}{\textbf{RQ4: What types of topics are mainly being discussed in the gender dysphoria and gender euphoria TikTok online space?}
}}
\end{center}

\begin{figure*}[t]
\centering
\includegraphics[width=\linewidth]{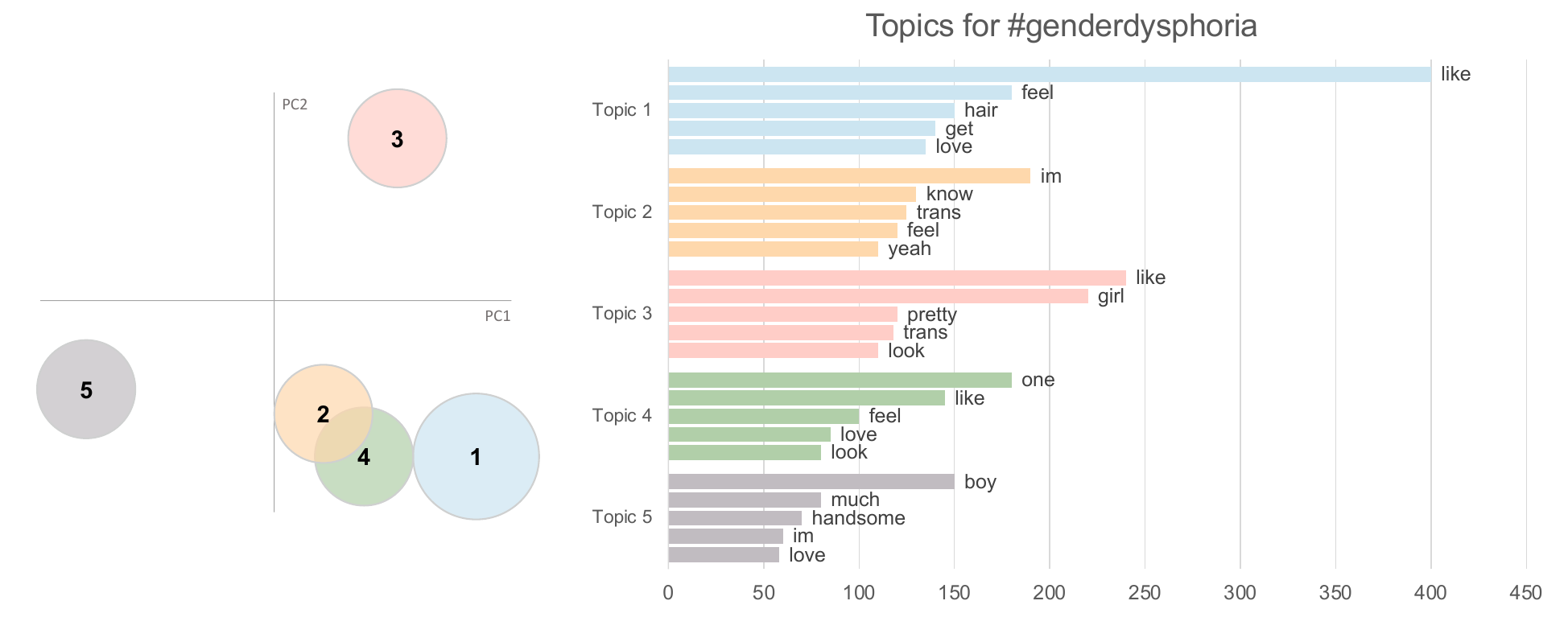}
\caption{LDA analysis for \#Genderdysphoria comments. For each topic, we show the top 5 words.}
\label{fig:lda_dys}
\end{figure*}

To answer \textbf{RQ4}, we used \textsc{Latent Dirichlet Allocation} (LDA), an unsupervised topic modeling method to obtain the most representative words in the comments we collected~\cite{blei2003latent}. We focus on 5 topics for each hashtag, visualizing the inter-topic distance in two dimensions, and illustrating the top-5 most relevant terms for each topic within the LDA graphs. 
Based on these rankings, we can compare how people express their views of gender euphoria and gender dysphoria and the terms they used in comment sections to describe those experiences.

% \subsubsection{LDA for \#Genderdysphoria}
\label{sec:lda_dys}
Figures~\ref{fig:lda_dys} illustrates our LDA analysis clusters of the comments with the hashtag of gender dysphoria. Clusters 1, 3, and 5 contained the most salient information. Cluster 3 is best represented by the term “girl’ referencing the more transfeminine experience of gender dysphoria (i.e. contains most references to and is, therefore, most different in its focus on that topic). This is also evidenced by the presence of the feminine gendered term “pretty”, alongside “girl” among the top 5 terms used. Just as in our word clouds relating to both gender dysphoria and gender euphoria (Figures~\ref{fig:wordCloud_cmmt}, \ref{fig:wordCloud_desc}), terms referencing the visual field were quite prominent. Further, the words “look” and “want”, relating to desire, are common in the comment section of Cluster 3.

Cluster 5, in contrast, illustrates comments largely represented by the term “boy” along with the masculine gendered term “handsome”, in the top five terms. Cluster 5, then, likely depicts the more transmasculine experience on TikTok. The term “want” is the sixth most commonly used term in these comments, but unlike Cluster 3, “look” is not among the most common words. Clusters 3 and 5 appear to be maximally distant in Figure 5, indicating that the topics centered in the comment sections of TikToks relating to gender dysphoria are very different for the transmasculine and transfeminine experiences.

Cluster 1 is equidistant from both Cluster 3 and Cluster 5 and focuses predominantly on the visual field in these transgender dysphoric experiences. Cluster 1 is most represented by the term “hair” in the comments, reinforcing the data in the word clouds shown in Figure~\ref{fig:wordCloud_cmmt_dys} and \ref{fig:wordCloud_desc_dys} about the importance of hair appearance for those experiencing gender dysphoria. Terms such as “like”, “feel”, “get”, and “love” all hint at a kind of personal emotive experience and desire in videos relating to gender dysphoria. Overall, these results support our findings about gender dysphoria (see Table~\ref{tab:sentiment}) from the sentiment analysis of comments: The terms used do not connote strong negativity, though the topic of TikTok is not a positive one.

\begin{figure*}[t]
\centering
\includegraphics[width=\linewidth]{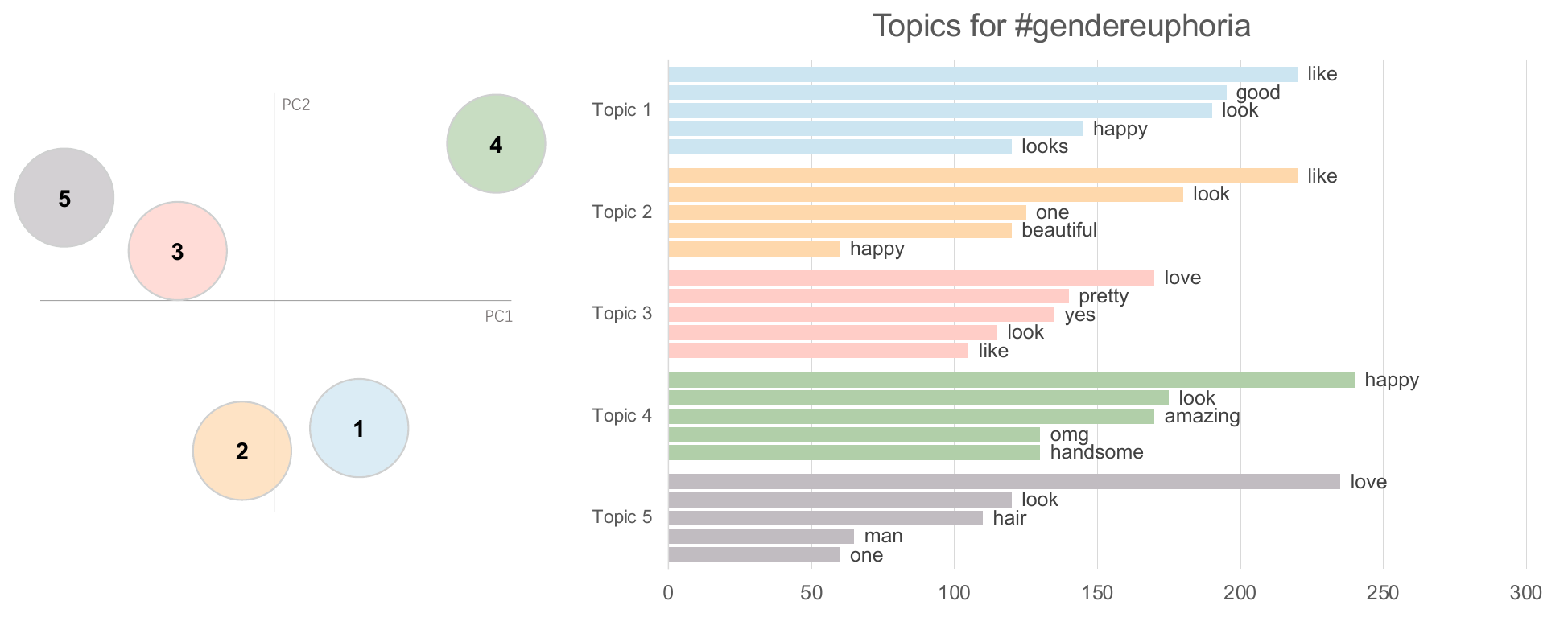}
\caption{LDA analysis for \#Gendereuphoria comments. For each topic, we show the top 5 words.}
\label{fig:lda_eu}
\end{figure*}

% \subsubsection{LDA for \#Gendereuphoria}
\label{sec:lda_eu}

Figure~\ref{fig:lda_eu} illustrates the clusters found by the LDA analysis of TikTok comments tagged gender euphoria. The 5 clusters focus on different positive terms, bolstering the results of our sentiment analysis, indicating an overall positive weight for comments pertaining to gender euphoria. 

Cluster 1 is most illustrated by the terms “like”, “good”, and “look”, as well as “happy”, indicating the importance of visuality and the joy visible to commenters or experienced by commenters watching these TikToks. Cluster 2’s most significant term is “beautiful”, and overlaps significantly with Cluster 1, which also centers words such as “like”, “look”, and “happy”. Both these two clusters demonstrate a strongly positive sentiment overall.

Clusters 3 and 5 sit in a different quadrant from Cluster 1 and 2, and likewise overlap significantly. Both Clusters 3 and 5 indicate that terms such as “love”, “look”, and “like” appear as the top five most used words in the comment sections analyzed. Cluster 3 also includes the term “pretty", whereas Cluster 5 includes terms such as “man” and “hair”, indicating that Cluster 3 likely represents comments more focused on transfeminine experiences of gender euphoria, in contrast to Cluster 5, which likely represents a more transmasculine experience. In addition to Cluster 5, Cluster 4 also seems to represent a transmasculine experience, with the term “handsome” being most used in its clustering of commented words. This could indicate, just as in our word clouds (Figures~\ref{fig:wordCloud_cmmt}, \ref{fig:wordCloud_desc}), a greater transmasculine presence in TikTok spaces relating to gender euphoria. Further, if Cluster 5 largely relates to comments focusing on transmasculine gender euphoria, it raises the question of whether the word “hair” is a less gender neutral term in practice within TikTok comments, and whether it tends to be associated, as our results indicate, with conversations about transmasculinity. 

Cluster 4 dovetails with other clusters in terms of its overall positive word associations, and includes words such as “look”, “happy”, and “amazing”. There were far fewer overarching differences in the clusters associated with gender euphoria than with those associated with gender dysphoria. Gender dysphoria LDA analysis delineated clear transmasculine and transfeminine clusters of comments, whereas the differences in comment sections relating to gender euphoria were much less salient. Though gender dysphoria significantly differentiates between transmasculine and transfeminine experiences, the terms used to describe gender euphoria appear to have wide resonance across transgender lives.

\section{Discussion}
\label{sec:disscuss}

In answering \textbf{RQ1} - \textbf{RQ4}, the synthesis of our results across methodologies was quite telling. As expected, the two gender experiences, gender dysphoria and gender euphoria, are characterized by different terms in the description section of TikTok videos. Our Spearman’s $k$ Correlation (Figure~\ref{fig:spearman_desc}) analyzed the description section of TikToks under the hashtags for \#genderdysphoria and \#gendereuphoria and indicated that when people describe their own experiences of gender dysphoria and gender euphoria, they use different terms to do so. This analysis found that these two experiences are differentiated by TikTok users and was expected based on clinical research into gender dysphoria and gender euphoria~\cite{galupo2020every,jacobsen2022counts,jacobsen2022moving,johnson2019rejecting,konnelly2022transmedicalism,mackinnon2021examining,pulice2020certain}. 

Our data support gender diverse content creators’ claims on TikTok that these spaces are very uplifting, and indicates the positive impacts of community building for gender diverse individuals online\cite{singh2013transgender}. Because commenters are primarily able to comment on the visual field of a video, which is a limited foray into the personal world of another person, the comments tended to focus on optical materialities, evidenced by the results of our word clouds (Figure~\ref{fig:wordCloud_cmmt}) and LDA analysis (Figure~\ref{fig:lda_dys} and \ref{fig:lda_eu}) evaluating comment sections which showed consistent centering of terms such as “look”. Our semantic network analysis indicated that “look” was a term used most commonly in comment sections in regards to gender euphoria, whereas in comment sections in regards to gender dysphoria, terms such as “feel” tended to be used more often. Words such as “love” and “like” also appeared consistently in results focused on comment sections, including word clouds (Figure~\ref{fig:wordCloud_cmmt} and \ref{fig:wordCloud_desc}), semantic network analysis (Figure~\ref{fig:semantic_cmmt}), and LDA analysis (Figure~\ref{fig:lda_dys} and \ref{fig:lda_eu}) of both gender dysphoria and gender euphoria. Our sentiment analysis of hashtags and comments may point to why: commenters were consistently more positive than content creators, and the presence of positive terms such as “love” and “like” in comments may indicate commenters’ attempts to uplift or otherwise positively influence transgender content creators talking about their gender dysphoria or gender euphoria. Also, our Spearman’s $k$ correlation analysis of TikTok comments (Figure~\ref{fig:spearman_cmmt}) found a wide difference between comment sections for videos about gender euphoria and gender dysphoria, such that commenters were significantly more positive about gender euphoria, but not significantly negative about gender dysphoria. This finding is supported by our sentiment analysis on comments regarding gender dysphoria, in which commenters were also significantly more positive than content creators.

Further, our findings reflect the experience of some transfeminine voices, which argue that transmasculine perspectives on TikTok tend to dominate online conversations. Our word clouds (Figure~\ref{fig:wordCloud_cmmt} and \ref{fig:wordCloud_desc}), semantic network analysis (Figure~\ref{fig:semantic_cmmt}), and LDA analysis (Figure~\ref{fig:lda_dys} and \ref{fig:lda_eu}) all indicate that transmasculine voices may dominate TikTok spaces devoted to gender euphoria and gender dysphoria based on the increased presence of terms and the heightened influence of clusters relating to the transmasculine gender experience. However, these findings are not conclusive. Further research using these tools are needed to determine whether this is a fact of our sampling or a fact of the TikTok online space. 

Finally, our research indicates that while gender dysphoria may be more delineated by transmasculine versus transfeminine gender trajectories and experiences, gender euphoria may be a far more broad-based experience across transgender lives. Our LDA analysis further indicates that in comment sections the terms transmasculine and transfeminine people use to describe their experiences of gender dysphoria are more divergent than the words used to describe experiences of gender euphoria. Clusters from the gender euphoria LDA analysis (Figure~\ref{fig:lda_eu}) showed more consistent overlap in the graphs as well as more similar word implications than that of the LDA analysis focused on gender dysphoria (Figure~\ref{fig:lda_dys}). Also, the LDA analyses focused on gender dysphoria clearly delineated transmasculine and transfeminine clusters and terms, which was a pattern absent in the LDA analysis focused on gender euphoria. This is a fascinating finding that deserves greater research.

\section{Related Work}

There has been significant research into how TikTok operates and how knowledge dissemination tends to operate on social media sites ~\cite{fiallos2021tiktok,garcia2022tiktok,hayes2020making,kong2021tiktok}. ~\cite{tao2020mgat} have done model research on how user preferences function on TikTok, for example. ~\cite{zhuang2023makes} have measured the influence of user-generated content and its helpfulness for viewers, and ~\cite{li2022lifecycle} have researched how rumor functions across these online community spaces. The importance of knowledge dissemination is critical, as ~\cite{langley2018collective} have analyzed in the context of community health interventions and improving the lives of patients. Social media sites can be a powerful force in support of knowledge sharing ~\cite{castillo2021effectiveness,chan2020social,huang2019communication,khan2021social}, including for the sharing of scientific information and medical publications ~\cite{johannsson2020dissemination}. ~\cite{mackinnon2021examining}, in an important study, has analyzed how content about transgender communities can move across the app, but was limited to the analytics of MacKinnon’s TikTok account itself, and not a wider net of analysis across TikTok using different approaches. Our research complements this approach by analyzing content about trans people broadly across TikTok.

Social media, particularly TikTok and YouTube, have become important sites of community building and interaction among LGBTQ individuals~\cite{fredenburg2020youtube,lovelock2017every,miller2019youtube,rodriguez2022lgbtq}. Social media sites are driven by algorithms that filter, manage, and direct content from creators in positive and negative ways, which users can resist or at other times cannot ~\cite{gillespie2015platforms,grandinetti2021examining,grandinetti2022affective,karizat2021algorithmic,siles2022learning,simpson2022tame}(Gillespie 2015). This is especially the case on TikTok, where every user has access to a highly personalized For You Page on which in-app algorithms determine what content will appear. Thus, analyzing these connections using different methodologies is a critical next step in research about LGBT+ communities online, particularly transgender communities. There is a dearth of such approaches, although ~\cite{DBLP:journals/corr/abs-2010-13062} study on the sentiment analysis of transgender people’s social media posts about mental health conditions is extremely relevant to such a research agenda. Studies like ~\cite{DBLP:journals/corr/abs-2010-13062} and this research demonstrates the potentially surprising associations of trans communities in virtual spaces and are critical to supporting this growing minority community.

Previous studies on gender dysphoria and gender euphoria specifically have tended to utilize social media as a method for recruitment~\cite{beischel2022little,jacobsen2022counts,jacobsen2022moving,johnson2019rejecting,konnelly2022transmedicalism,mackinnon2021examining,galupo2020every,pulice2020certain}, but have relied on interview methods, which are limited in terms of the population of participants that can be included. Topical and sentiment methodological analysis allows for a wider engagement with different users about personal experiences, such as gender dysphoria, and allows the data on that topic to arise not from specific questions posed, which may affect the type of data collected, but rather from the discourse already occurring between trans people. Our research intervenes into the study of gender dysphoria using topic analysis and illustrate how trans communities discuss gender dysphoria, gender euphoria, and the expected and potentially surprising ways in which these conversations align.

\section{Conclusion and Future Work}
\label{sec:conclusion}

This study highlights the effectiveness of topical and sentiment analysis in understanding interactions among transgender communities on TikTok. Unlike previous research focusing on user-based analytics~\cite{mackinnon2021examining}, our analysis of video content itself provides new insights into community formation and engagement.

Our methods reveal interaction patterns, content types, and terminology that are critical for researchers studying transgender and other marginalized communities on TikTok. These findings offer a framework for understanding how these communities discuss and interact with various topics.

Future research could investigate the extent to which gender-diverse communities on TikTok support and affirm creators. Key areas of interest include the support received by different creators, the expression of gender euphoria versus dysphoria, and the potential dominance of transmasculine voices. Longitudinal sentiment analysis, comparing comments with content hashtags and descriptions, could provide deeper insights into these dynamics and self-perception in virtual environments~\cite{DBLP:journals/pacmhci/LeeMEH22}.

\begin{credits}
\subsubsection{\ackname} This material is based upon work supported by NSF (SaTC-2241068 and IIS-2339198), and a Cisco Research Award.

% \subsubsection{\discintname}
% It is now necessary to declare any competing interests or to specifically
% state that the authors have no competing interests. Please place the
% statement with a bold run-in heading in small font size beneath the
% (optional) acknowledgments,
% for example: The authors have no competing interests to declare that are
% relevant to the content of this article. Or: Author A has received research
% grants from Company W. Author B has received a speaker honorarium from
% Company X and owns stock in Company Y. Author C is a member of committee Z.
\end{credits}

\bibliographystyle{splncs04}
\bibliography{reference}

\end{document}